\documentclass[reprint]{revtex4-1}
\usepackage{graphicx}
\usepackage{latexsym}
\usepackage{hyperref}
\usepackage{amsmath,amsthm,amssymb}
\usepackage{natbib}
\usepackage{epsfig}
\usepackage{subfigure}
\usepackage{epstopdf}

\begin{document}
\title{Evolution of cooperation in spatial traveler's dilemma game}

\author{Rong-Hua Li}
\author{Jeffrey Xu Yu}
\affiliation{Department of Systems Engineering \& Engineering
Management, The Chinese University of Hong Kong - Hong Kong, China}
\author{Jiyuan Lin}
\affiliation{Institute of Computing Technology, Chinese Academy of
Sciences, Beijing, People¡¯s Republic of China}


\begin{abstract}
Traveler's dilemma (TD) is one of social dilemmas which has been well studied in the economics community, but it is attracted little attention in the physics community. The TD game is a two-person game. Each player can select an integer value between $R$ and $M$ ($R < M$) as a pure strategy. If both of them select the same value, the payoff to them will be that value. If the players select different values, say $i$ and $j$ ($R \le i < j \le M$), then the payoff to the player who chooses the small value will be $i+R$ and the payoff to the other player will be $i-R$. We term the player who selects a large value as the cooperator, and the one who chooses a small value as the defector. The reason is that if both of them select large values, it will result in a large total payoff. The Nash equilibrium of the TD game is to choose the smallest value $R$. However, in previous behavioral studies, players in TD game typically select values that are much larger than $R$, and the average selected value exhibits an inverse relationship with $R$. To explain such anomalous behavior, in this paper, we study the evolution of cooperation in spatial traveler's dilemma game where the players are located on a square lattice and each player plays TD games with his neighbors. Players in our model can adopt their neighbors' strategies following two standard models of spatial game dynamics. Monte-Carlo simulation is applied to our model, and the results show that the cooperation level of the system, which is proportional to the average value of the strategies, decreases with increasing $R$ until $R$ is greater than the threshold where cooperation vanishes. Our findings indicate that spatial reciprocity promotes the evolution of cooperation in TD game and the spatial TD game model can interpret the anomalous behavior observed in previous behavioral experiments.

%
%

%
%

\end{abstract}

\pacs{87.23.Kg,89.65.-s,87.10.Rt}
\pagestyle{empty} \maketitle

\section{Introduction}
\label{intro}
Cooperation is ubiquitous in biological and social systems \cite{06evolcooperation, *97bookanimalcoopera, *06sciencefiverule}. In general, cooperation is expensive, which leads to the so-called social dilemma. For a social dilemma, a group of individuals can achieve the maximal payoff by cooperation, but individuals perform best by acting in their own interests. Understanding the origins of cooperation in a group of unrelated and self-interested individuals is a central problem in biological, social, and physical science \cite{06sciencefiverule}. The evolutionary game theory is an elegant framework to study such problem \cite{97bookegametheory, *98bookegamepd}. The widely studied evolutionary game models include evolutionary prisoner's dilemma game \cite{92naturespatialgame, 06prepreferselectpd, 08prediversityspdg}, evolutionary snowdrift game \cite{07phyreportegongraphs}, and evolutionary public goods game \cite{07prenoisepgg, 08naturediversitypgg, 09pretopologyspgg}. Both the prisoner's dilemma (PD) game and the snowdrift (SD) game are the two-person game where the players can choose either cooperation or defection. The only difference between the PD and SD games is the payoff matrix \cite{07phyreportegongraphs}. The public goods game, however, is an N-person game where the players can choose either to contribute to the public good (cooperation) or to contribute nothing (defection). All three types of games create social dilemma, and have been extensively studied in recent years \cite{07phyreportegongraphs, 09prethreestrategiespdg, 09prediversitypdg, 10epjbnonuniformspgg, 10precoevolvtimespd, 12precontinusstrategy, 10predefectorpgg, 10precriticalmasspgg, 11pregroupsizespgg, 11punishmentspgg, 12precondstrategyspgg}.

In this paper, we consider the traveler's dilemma (TD) game which has received extensive attention in the economic society but has attracted little attention in the physics community so far. Similar to the PD game, the TD game is also a two-person game which is proposed by Basu \cite{94td}. We give a brief description of the TD game as follows: assume that two travelers have identical souvenirs and both of which have been lost by the airline. The two travelers come back to their airline to ask for compensation. The airline representative does not know the accurate price of the souvenirs, but he knows that the price falls within an interval $[R, M]$. Therefore, the airline representative asks the two travelers to write down the value from $R$ to $M$ separately. If both travelers claim the same value, then the airline will compensate both with that amount. However, if they declare different values, the airline representative will assume that the lower value is more accurate. Therefore, the representative pays the traveler who claims the lower value that amount plus a bonus of $R$ for his honesty, and gives the other traveler the lower value minus $R$ for penalty. For example, if one traveler declares that the price of the souvenir is 20 while the other traveler declares that its price is 30. Suppose $R=2$, then the first traveler will receive 22 while the other will get 18. Following \cite{12jtbtd}, we assume that both travelers declare an integer number, and both $R$ and $M$ are integer number. To create a social dilemma, we restrict $R>1$, similar restriction has been done in \cite{12jtbtd}.

By the classical game theory, the Nash equilibrium of the TD game is that both travelers claim the minimal number $R$ \cite{94td}. Clearly, the maximal total payoff of the travelers is $2M$ by both declaring the maximal value $M$. As a result, the TD game yields a social dilemma. Many previous experimental studies found that the players' behavior significantly deviated from the prediction of the classical game theory. Capra et al.\ \cite{99aertd} found that there exists an inverse relationship between $R$ and the average claim. That is to say, for a small $R$, the average claim could be a large value. Subsequently, Goeree and Holt \cite{99pnastd} presented a learning framework to interpret such anomalous behavior. More recently, Manapat et al.\ \cite{12jtbtd} proposed a stochastic evolutionary framework to explain the cooperation behavior observed in TD game. Specifically, they studied stochastic evolutionary dynamics in finite populations with varying selection and mutation rate parameters, and their theoretical results confirmed the observed cooperation behavior. In this paper, we study TD game on a square lattice by adopting the standard spatial game model. Using Monte-Carlo simulation, we find that the observed cooperation behavior in our system is consistent with the previous experimental observations. Furthermore, we also present an analysis on an ideal model where the players can only select two pure strategies ($R$ and $M$) to explain the observed phenomenon which further confirms our results. Our findings indicate that the spatial reciprocity can facilitate the evolution of cooperation in TD game, and thereby the spatial TD game model can be used to interpret the observed cooperation behavior in TD game.

The outline of the rest of the paper is as follows. First, in Section~\ref{sec:model}, we introduce the spatial TD game model, defining the payoff matrix and describing strategy adoption rules. Second, our simulation and analytic results are presented in Section~\ref{sec:results}. Finally, we conclude this work and point out some future directions in Section~\ref{sec:concld}.

\section{Model}\label{sec:model}
The TD game is a two-person game with multiple strategies. In TD game, each player selects a pure strategy from a discrete strategy space including $M-R+1$ strategies. For convenience, we label these strategies as $R, \cdots, M$, where $1 < R < M$. Without loss of generality, we set $M=100$, and similar setting has been considered in \cite{12jtbtd}. The payoff, denoted by $A_{ij}$, for a traveler claiming a value $i$ (strategy $i$) when the other declaring a value $j$ (strategy $j$), is given by
\begin{equation}
  \label{eq:payoff}
  A_{ij}  = \left\{ \begin{array}{l}
 i, \quad \quad \quad if \; i = j \\
 i + R, \quad if \; i < j \\
 j - R, \quad if \; i > j \\
 \end{array}. \right.
\end{equation}
In the above TD game, the Nash equilibrium is to choose the minimal value $R$ \cite{94td}. Similar to the prisoner's dilemma game, in TD game, defection (claiming a low value) will dominate cooperation (claiming a high value). In many previous behavioral studies \cite{99aertd, 99pnastd}, however, the researchers found that the players in TD game tended to select a much higher value than the minimal value. In this paper, we examine the impact of spatial structure in TD game. More specifically, we study evolutionary TD game in finite structured population where each player is located in a site of a square lattice with periodic boundary conditions. In our model, each player plays TD game with their nearest neighbors, and the total payoff of a certain player is the sum over all the payoffs gained by playing TD game with his neighbors. Following the standard spatial game model \cite{92naturespatialgame, 98pdgamesq}, a randomly chosen player $u$ can revise his strategy by adopting a strategy from his neighbors' strategies. We consider two strategy adaption rules. The first one is the deterministic strategy-adoption rule where the player always updates his strategy based on his payoff and his neighbors' payoffs. Specifically, under this rule, if the payoff of the player is smaller than the maximal payoff of his neighbors, the player adopts the strategy of his neighbor who has the maximal payoff, otherwise his strategy is unchanged. Similar deterministic strategy-adoption rule has been used for studying spatial prisoner's dilemma game \cite{92naturespatialgame, 06bookevodynamics}. The second one is the random strategy-adoption rule. In this rule, a player $u$ randomly selects one of his neighbor $v$ and adopts the strategy of player $v$ with the probability
\begin{equation}
  \label{eq:updateprob}W(s_u  \to s_v ) = \frac{1}{{1 + \exp [(P_u  - P_v )/\tau]}},
\end{equation}
where $s_u$ denotes the strategy of player $u$, and $\tau$ denotes the noise parameter modeling the uncertainty caused by strategy adoption. As explained in many previous studies \cite{98pdgamesq,09pretopologyspgg,10precriticalmasspgg}, for any finite positive $\tau$, better performing strategies are easier adopted and poor performing strategies are selected with a very small probability. At $\tau \to 0$ limit, the strategy adoption is nearly deterministic where the players will always select the better strategies, while at $\tau \to \infty$ limit, the strategy adoption is random.

We apply Monte-Carlo simulation to above spatial TD game. The size of the square lattice in our simulation is $100 \times 100$. We also perform simulation on a large-sized lattice (eg. $500 \times 500$), and no significant differences are observed. Initially, each player on site $u$ is randomly designated a strategy from $R$ to $M$ with equal probability, i.e., $1/(M-R+1)$. At each Monte-Carlo step, for the deterministic strategy-adoption rule, each player revises his strategy based on his payoff and his neighbor's payoff as described above. For the random strategy-adoption rule, each player randomly selects one of his neighbors $j$ and adopts the strategy of player $j$ according to the probability described in eq.~(\ref{eq:updateprob}). In all the simulations, we synchronously update the players' strategies. To measure the cooperation level of a system, we define a quantity denoted by $\rho_c$ as the normalized difference between the average value of all the players' strategies and the minimal value of strategy ($R$). More formally, $\rho_c$ is given by
\begin{equation}
  \label{eq:cooperlevel}
  \rho _c  = \frac{{\sum\nolimits_u {s_u } /n - R}}{M - R}.
\end{equation}
Clearly, $\rho _c$ is proportional to the average claim over all the players, and the value of $\rho _c$ falls within a range of [0, 1]. $\rho _c =0$ denotes that all the players declare the minimal value $R$ in which the system has the lowest cooperation level, and $\rho _c = 1$ denotes that all the players declare the maximal value $M$ where the system has the highest cooperation level. To ensure accuracy, we run 10,500 Monte-Carlo simulation steps. $\rho_c$ is obtained by averaging over the last 500 Monte-Carlo steps. All the results presented below are the average results over 30 realizations of initial strategies.

\section{Results}\label{sec:results}
\begin{figure}
\begin{center}
  \includegraphics[width=\columnwidth]{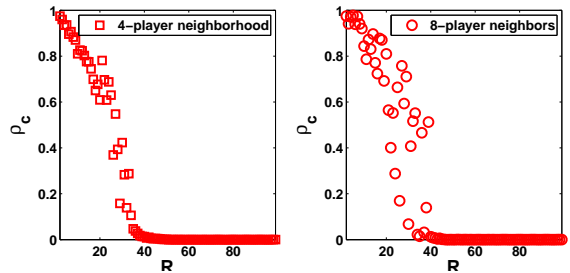}
\end{center}\vspace*{-2em}\caption[]{(Color online) Cooperation level $\rho_c$ as a function of the parameter $R$  on square lattices with 4-player neighborhood (left panel) and 8-player neighborhood (right panel) under the deterministic strategy-adoption rule.} \label{fig:fcr} \vspace*{-0.3cm}
\end{figure}

We start by reporting the results of the spatial TD game under deterministic strategy-adoption rule. Fig.~\ref{fig:fcr} depicts the simulation results for $\rho_c$ as a function of the parameter $R$ on two different square lattice models. For the square lattice with 4-player neighborhood (von Neumann neighborhood, left panel of Fig.~\ref{fig:fcr}), we can observe that (1) cooperation emerges given $R$ is smaller than the threshold $R_t$ ($R_t \approx 40$), and (2) cooperation level $\rho_c$ decreases monotonically with increasing $R$ until $R$ reaches $R_t$, where the cooperation level becomes 0. These results are consistent with the previous experimental observations in traditional TD game \cite{99aertd, 12jtbtd}, which show that there exists an inverse relationship between $R$ and the mean value claimed by the players in TD game. Further, the results suggest that our spatial TD game model can be used to interpret the anomalous behavior observed in traditional TD game. Similar results can be observed on the square lattice with 8-player neighborhood (Moore neighborhood). There are some minor differences in this model. First, the threshold value is slightly smaller than that of the previous lattice model. Second, there are certain points in the right panel of Fig.~\ref{fig:fcr} showing that $\rho_c$ does not decrease monotonically with increasing $R$, although the general results conform with those of the previous model. Similar to the results observed in traditional spatial game models (eg. spatial prisoner's dilemma game, spatial snowdrift game, and spatial public goods game), our findings indicate that spatial reciprocity also promotes cooperation in TD game. In the following, we will interpret the emergence of cooperation in spatial TD game and the observed phenomenon of the inverse relationship between $\rho_c$ and $R$ respectively.

\begin{figure}[t]
\begin{center}
 \subfigure[4-Player neighborhood, $R = 2$]{
     \includegraphics[width=\columnwidth, height=3cm]{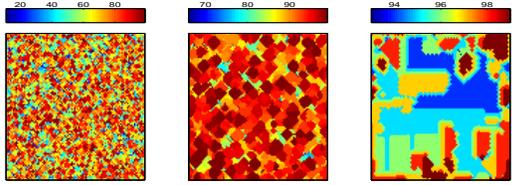}
 }
 \subfigure[8-Player neighborhood, $R = 2$]{
     \includegraphics[width=\columnwidth, height=3cm]{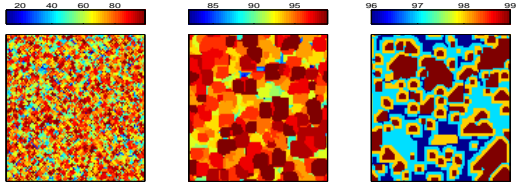}
  }
 \subfigure[4-Player neighborhood, $R = 10$]{
     \includegraphics[width=\columnwidth, height=3cm]{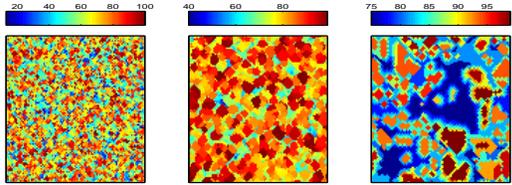}
 }
 \subfigure[8-Player neighborhood, $R = 10$]{
     \includegraphics[width=\columnwidth, height=3cm]{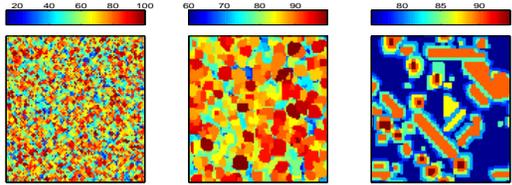}
  }
  \end{center}\vspace*{-2em}
\caption[]{(Color online) Characteristic snapshots describing the cooperation level of different $R$ and different square lattice models. (a) Snapshots of system's state on a square lattices with 4-player neighborhood given $R = 2$, (b) Snapshots of system's state on a square lattices with 8-player neighborhood given $R = 2$, (c) Snapshots of system's state on a square lattices with 4-player neighborhood given $R = 10$, and (d) Snapshots of system's state on a square lattices with 8-player neighborhood given $R = 10$. For $R=2$ ($R=10$), the initial state of both square lattices with 4-player neighborhood and 8-player neighborhood are identical random initial state. The middle and right snapshots of (a), (b), (c), and (d) are generated at 5th and 5000th Monte-Carlo iterations respectively.}
\label{fig:fcimage} \vspace*{-0.3cm}
\end{figure}

To reveal the potential mechanism behind the emergence of cooperation in spatial TD game, we can see the spatial patterns of the spatial TD game generated in our simulation. Figs.~\ref{fig:fcimage}(a-d) show a series of three characteristic snapshots taken at different times which describe the cooperation level of $R=2$ and $R=10$ on two different square lattice models respectively. Time evolution starts with a random initial state and ends in a stationary state (from the left snapshot to the right snapshot of Figs.~\ref{fig:fcimage}(a-d)). From the left snapshot to the right snapshot of Fig.~\ref{fig:fcimage}(a), we can observe that the cooperation level of the system increases with increasing iterations until the system goes to the stationary state. In addition, it can be seen that cooperators who declare the same large value will form scattered clusters (middle snapshot of Fig.~\ref{fig:fcimage}(a)), and such clusters spread out over the territory of defectors who declare small values. In the stationary state (right snapshot of Fig.~\ref{fig:fcimage}(a)), we can see that the strategy value becomes very large (close to $M$) and the number of different strategies becomes very small comparing with those in the initial state. Moreover, the players who declare the same large value will form a stationary cluster, and such stationary clusters can resist the invasion of the defectors.
 Similar results can be observed in Figs.~\ref{fig:fcimage}(b-d). These results indicate that the square lattice structure promotes the formation of clusters of cooperators, and thereby enhances the cooperation level of the system which further confirms that spatial reciprocity works well in TD game. In addition, by comparing the right snapshot of Fig.~\ref{fig:fcimage}(a) with the right snapshot of Fig.~\ref{fig:fcimage}(c), we can observe that the cooperation level of $R=2$ is clearly larger than the cooperation level of $R=10$. The reason is that, for $R=2$, the players with the same large strategy (nearly $M$) form a large cluster (see the right snapshot of Fig.~\ref{fig:fcimage}(a)), while for $R=10$, the size of such cluster is small. Furthermore, for $R=10$, there is a large territory occupied by the players who declare the same medium value (around 75). As a consequence, the cooperation level of $R=10$ is smaller than the cooperation level of $R=2$.




\begin{figure}[t]
\begin{center}
 \subfigure[t=1]{
     \includegraphics[width=0.2\columnwidth]{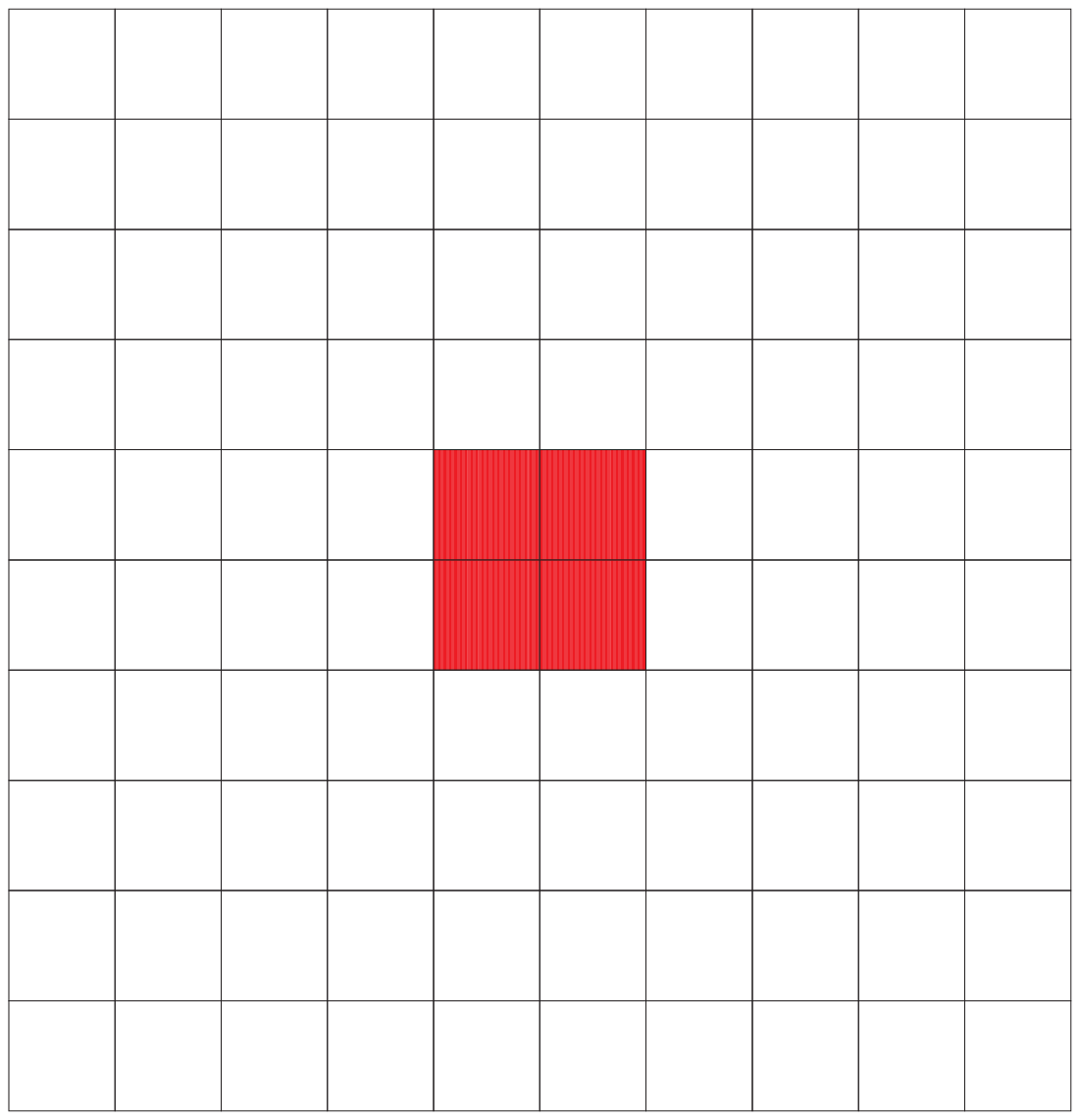}
 }
 \subfigure[t=2]{
     \includegraphics[width=0.2\columnwidth]{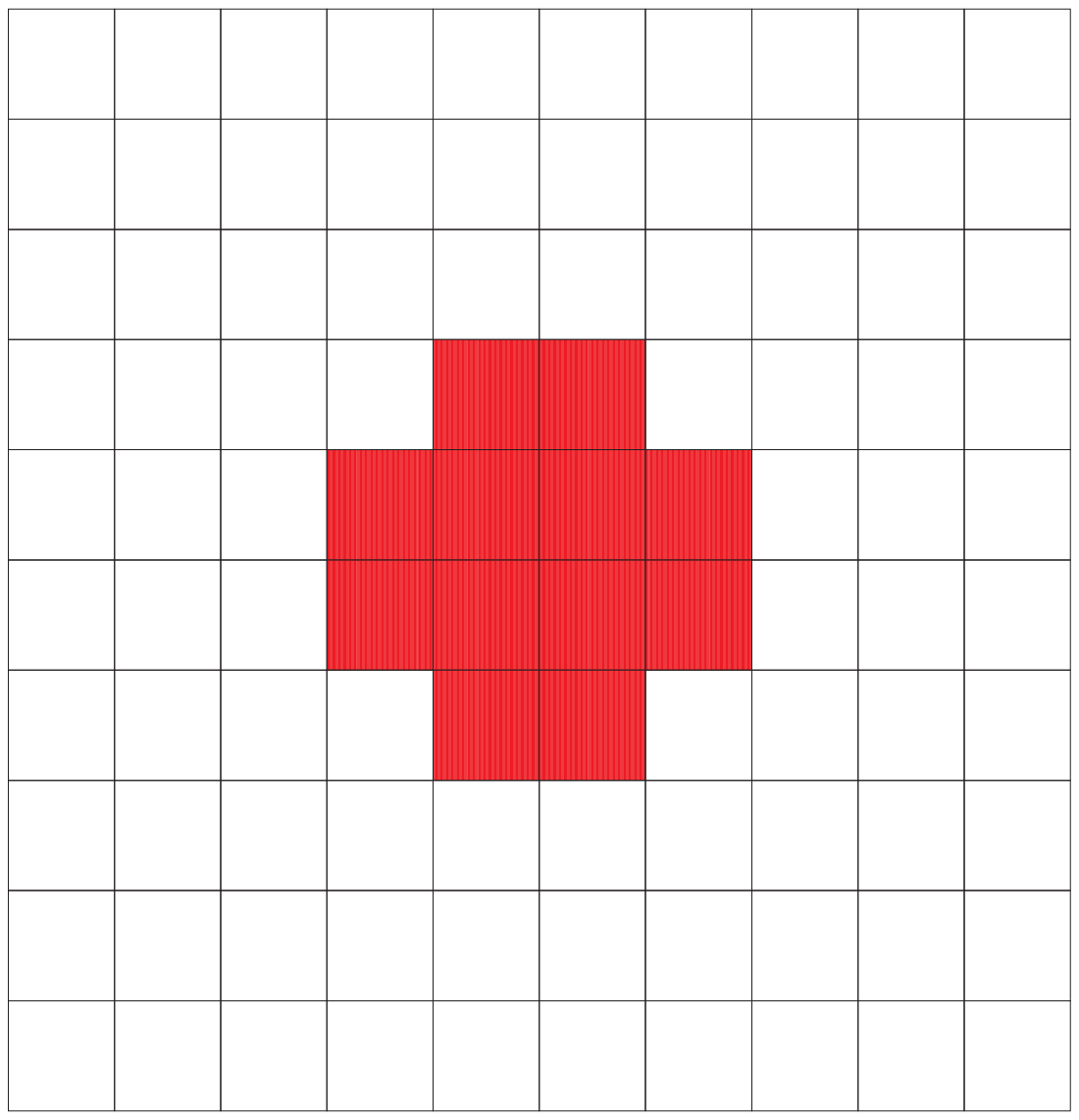}
  }
 \subfigure[t=3]{
     \includegraphics[width=0.2\columnwidth]{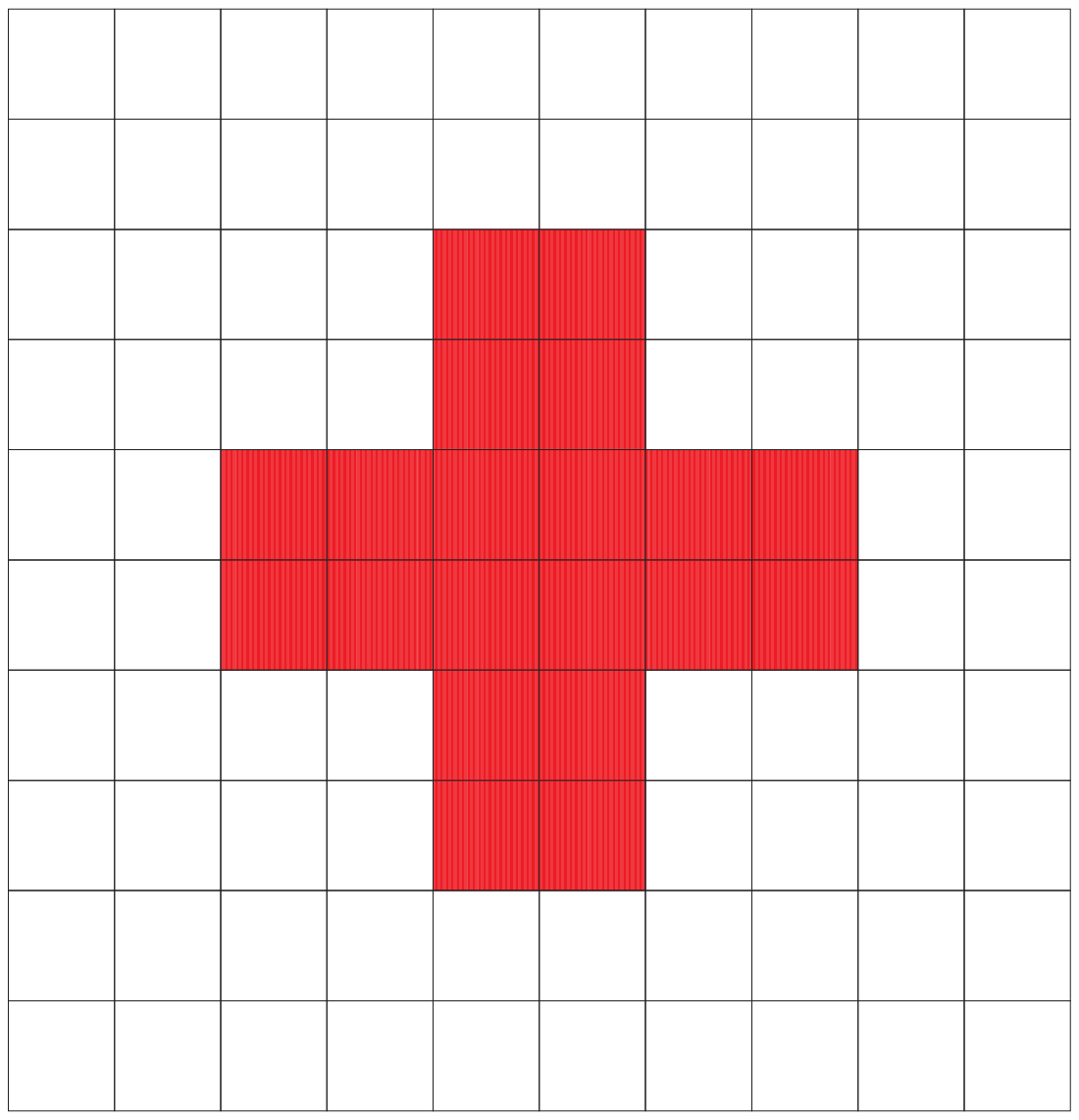}
  }
  \subfigure[t=4]{
     \includegraphics[width=0.2\columnwidth]{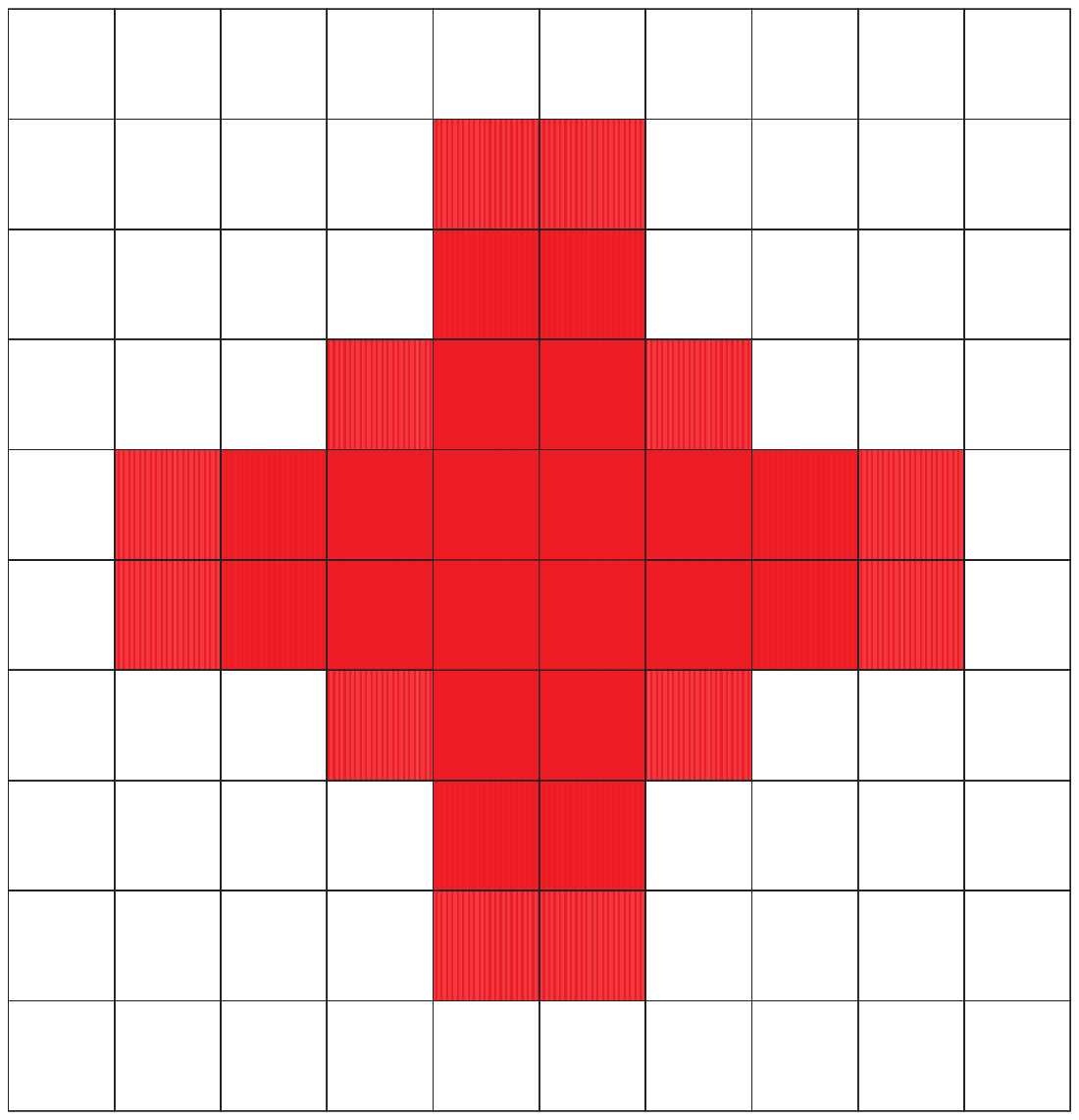}
  } \\
  \subfigure[t=5]{
     \includegraphics[width=0.2\columnwidth]{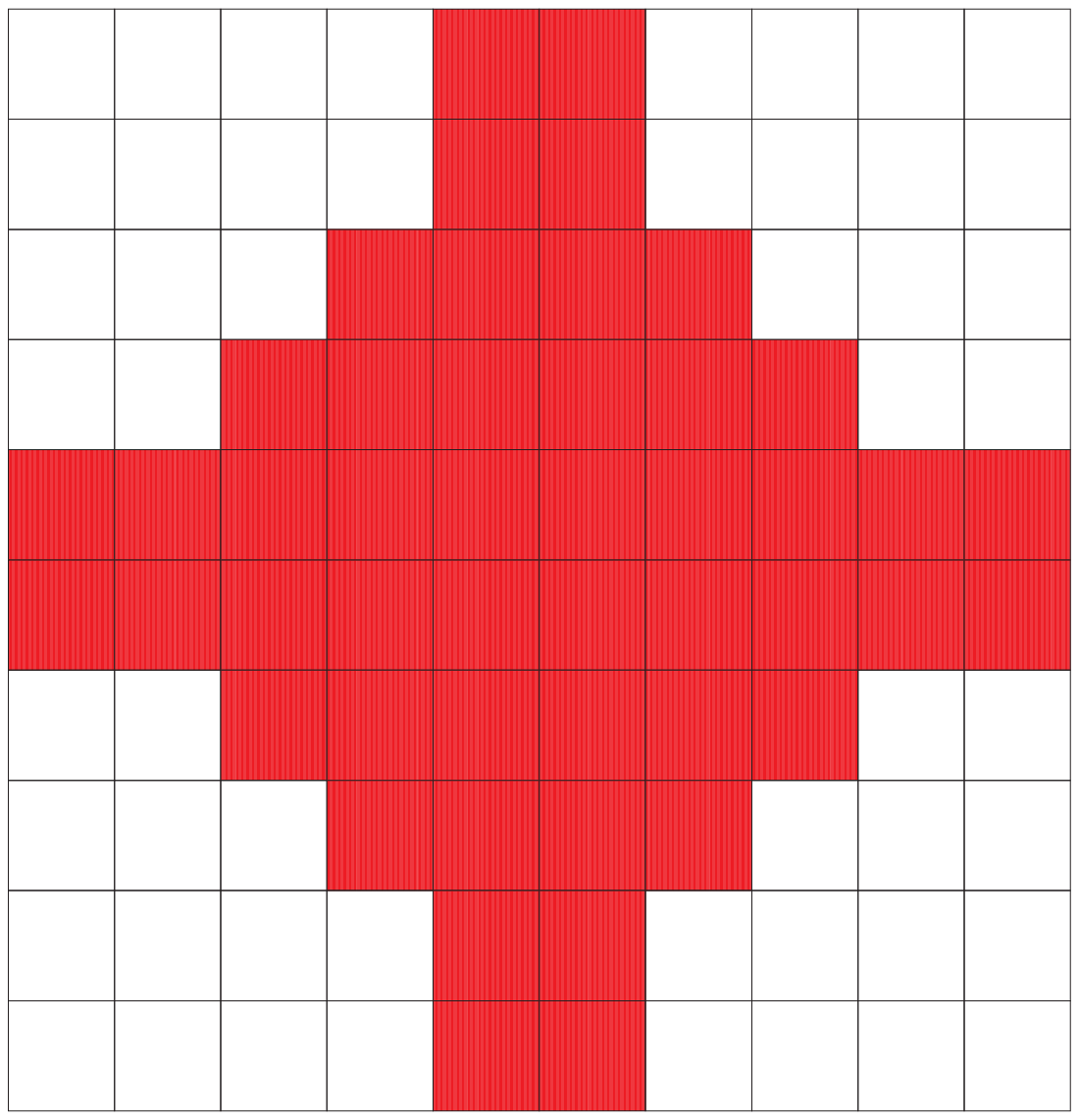}
 }
 \subfigure[t=6]{
     \includegraphics[width=0.2\columnwidth]{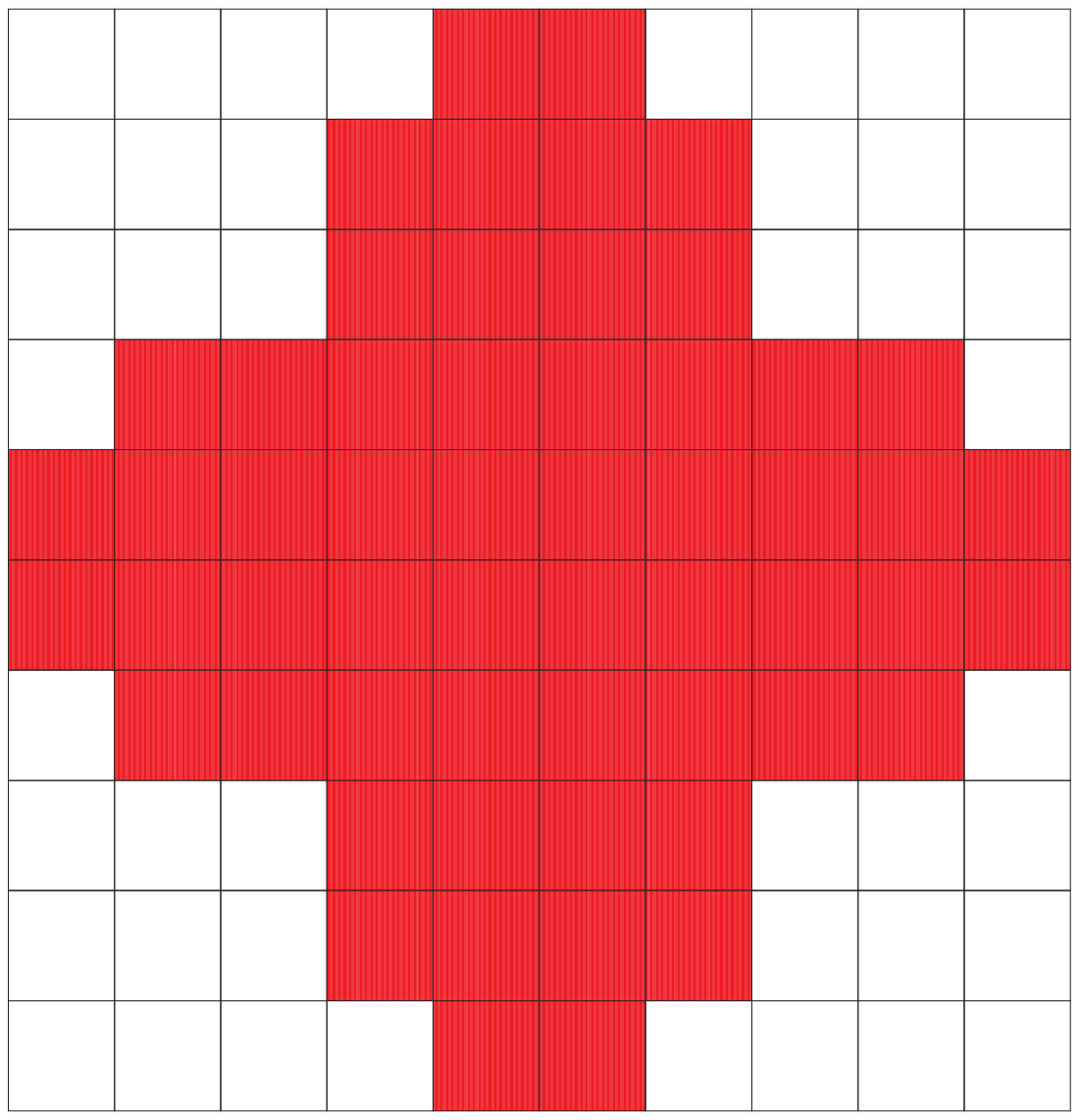}
  }
 \subfigure[t=8]{
     \includegraphics[width=0.2\columnwidth]{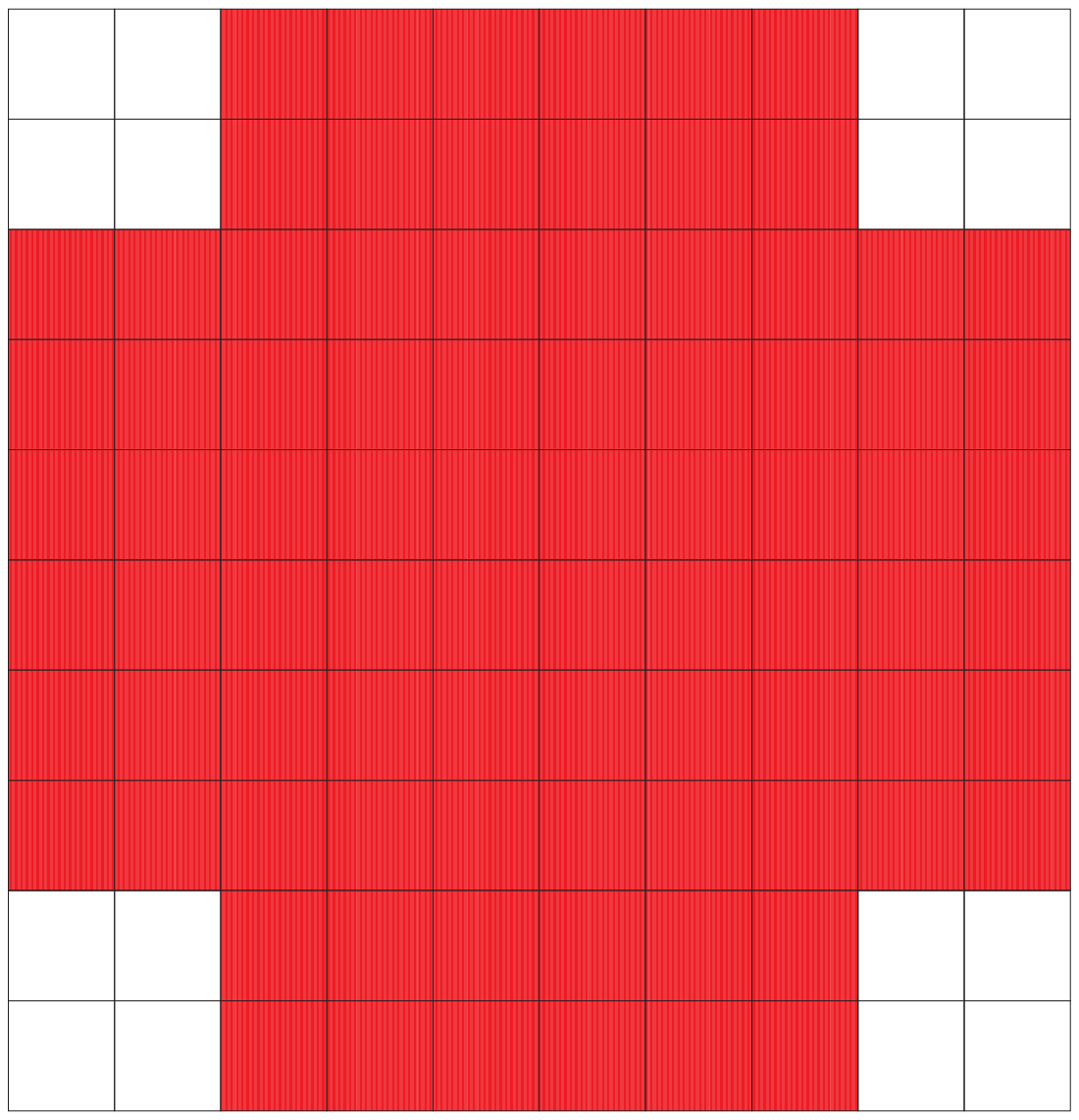}
  }
  \subfigure[t=10]{
     \includegraphics[width=0.2\columnwidth]{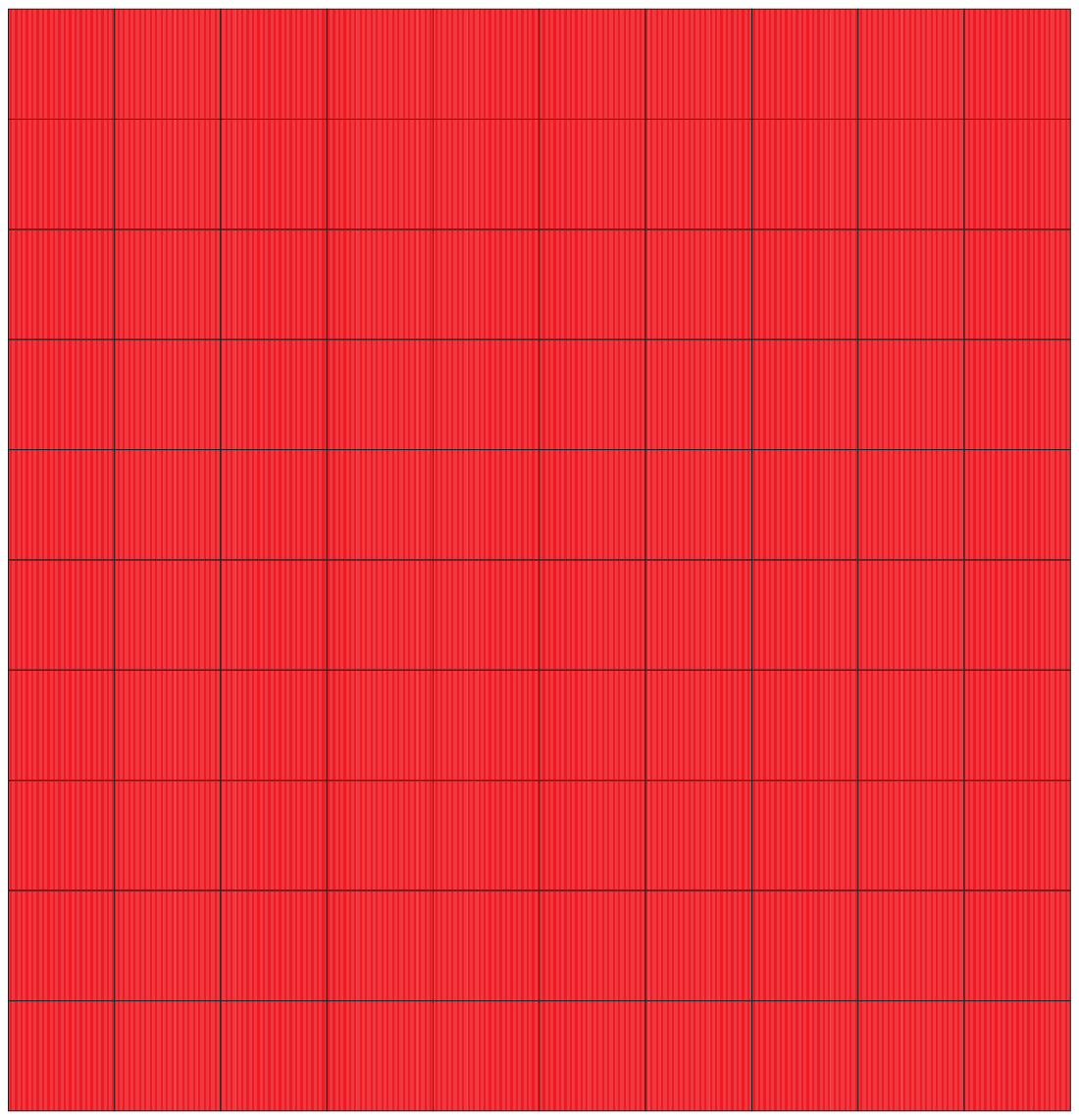}
  }
  \end{center}\vspace*{-2em}
\caption[]{(Color online) Time evolution of cooperator-invasion on a $10 \times 10$ square lattice with 4-player neighborhood. Initially, there are four cooperators who select strategy $M$ (the four red squares in figure (a)) and ninety-six defectors who choose strategy $R$ (the ninety-six blank squares in figure (a)). If $R < 2M/5$, then the cooperators occupy all the squares in the stationary state ($t=10$, figure(h)).}
\label{fig:c4invade} \vspace*{-0.3cm}
\end{figure}

\begin{figure}[t]
\begin{center}
 \subfigure[t=1]{
     \includegraphics[width=0.2\columnwidth]{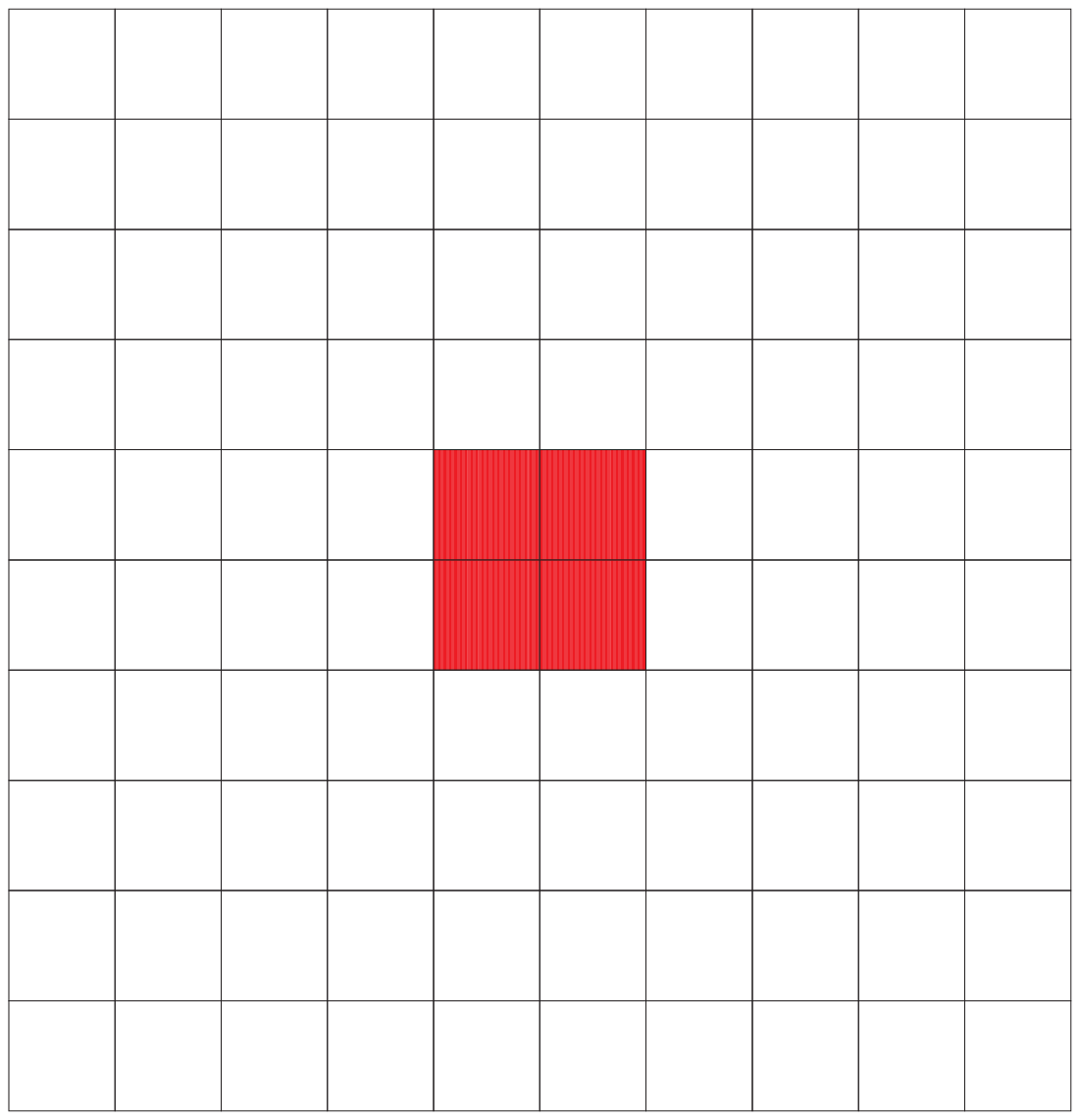}
 }
 \subfigure[t=2]{
     \includegraphics[width=0.2\columnwidth]{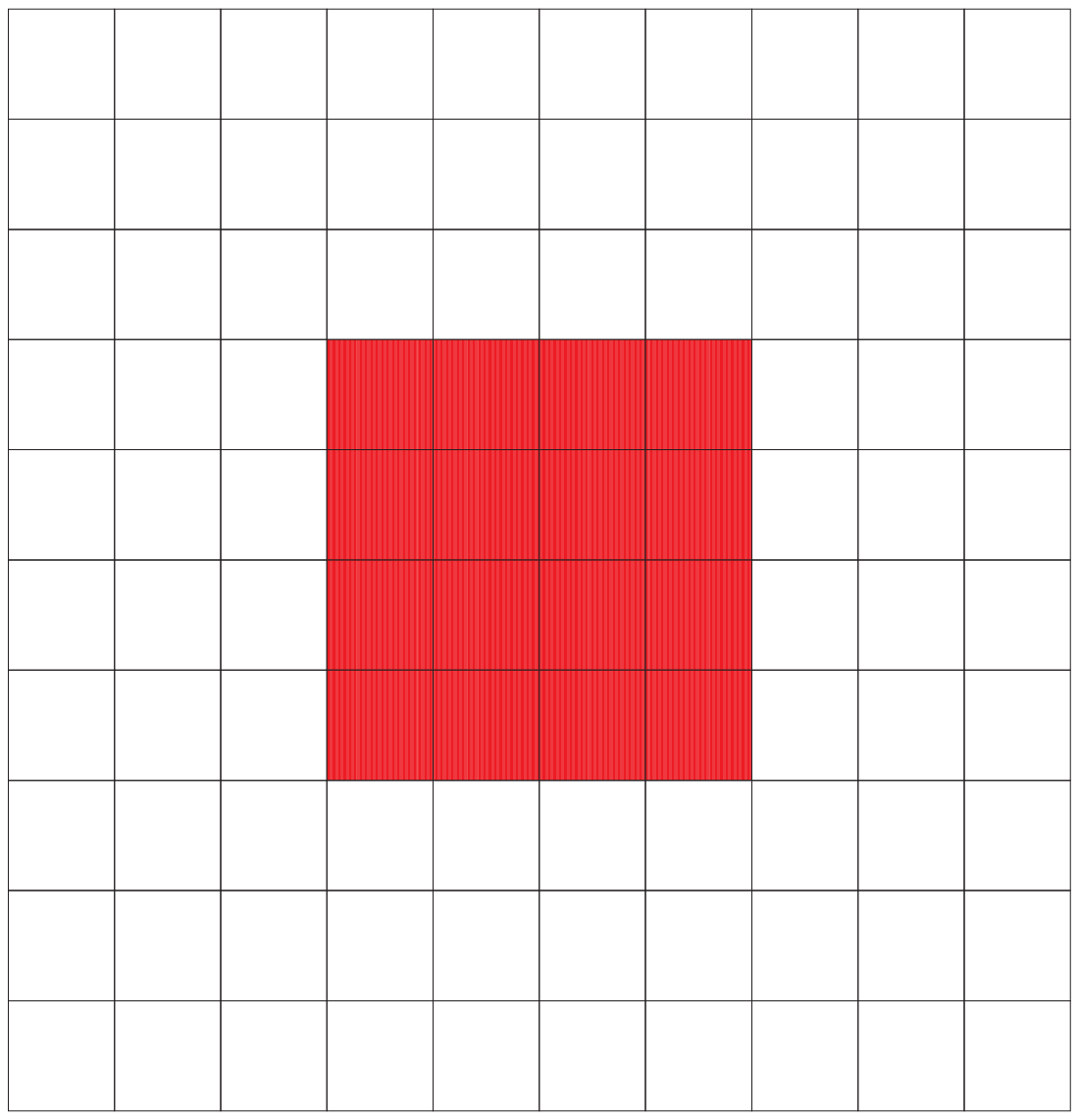}
  }
 \subfigure[t=3]{
     \includegraphics[width=0.2\columnwidth]{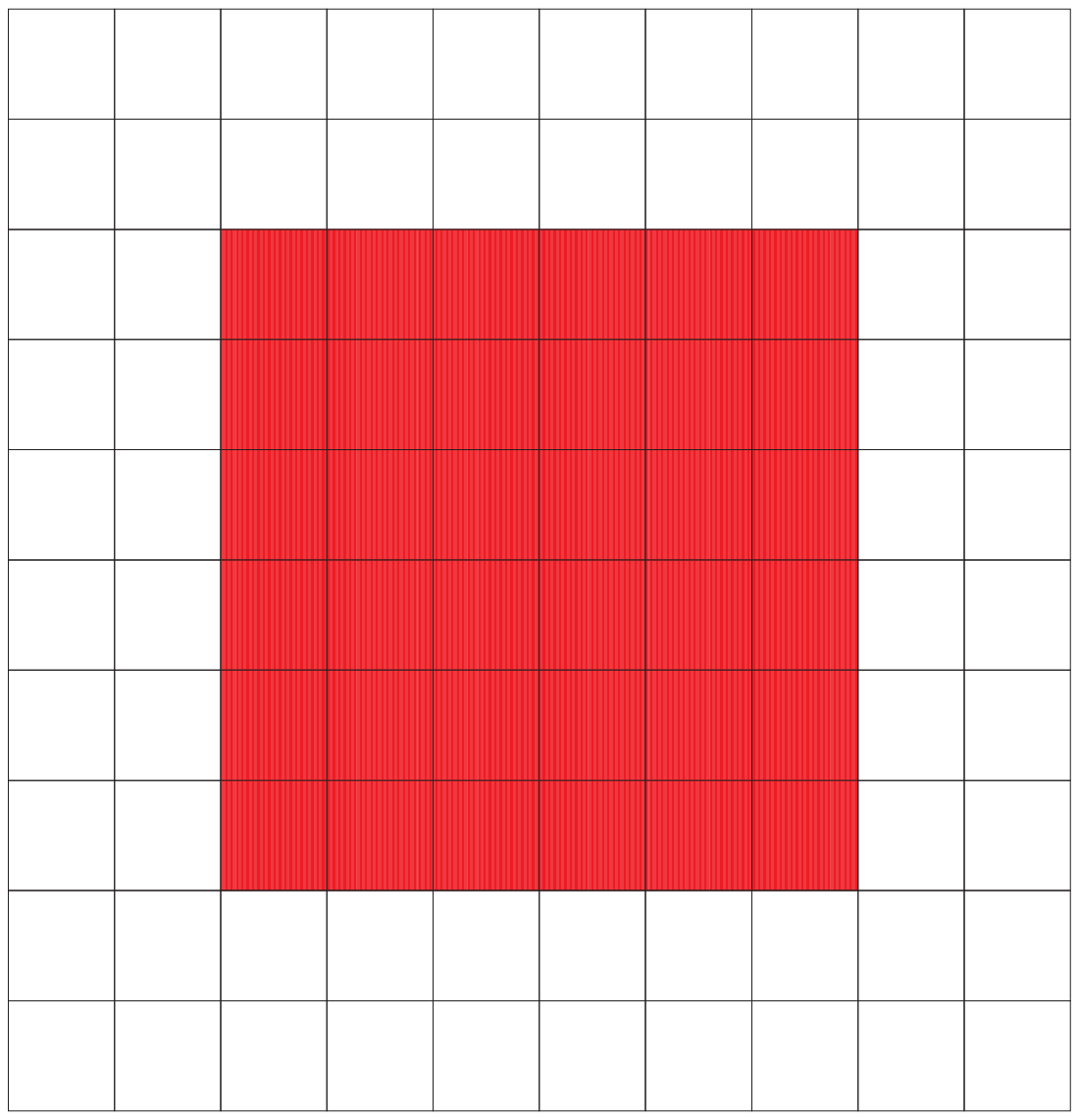}
  }
  \subfigure[t=5]{
     \includegraphics[width=0.2\columnwidth]{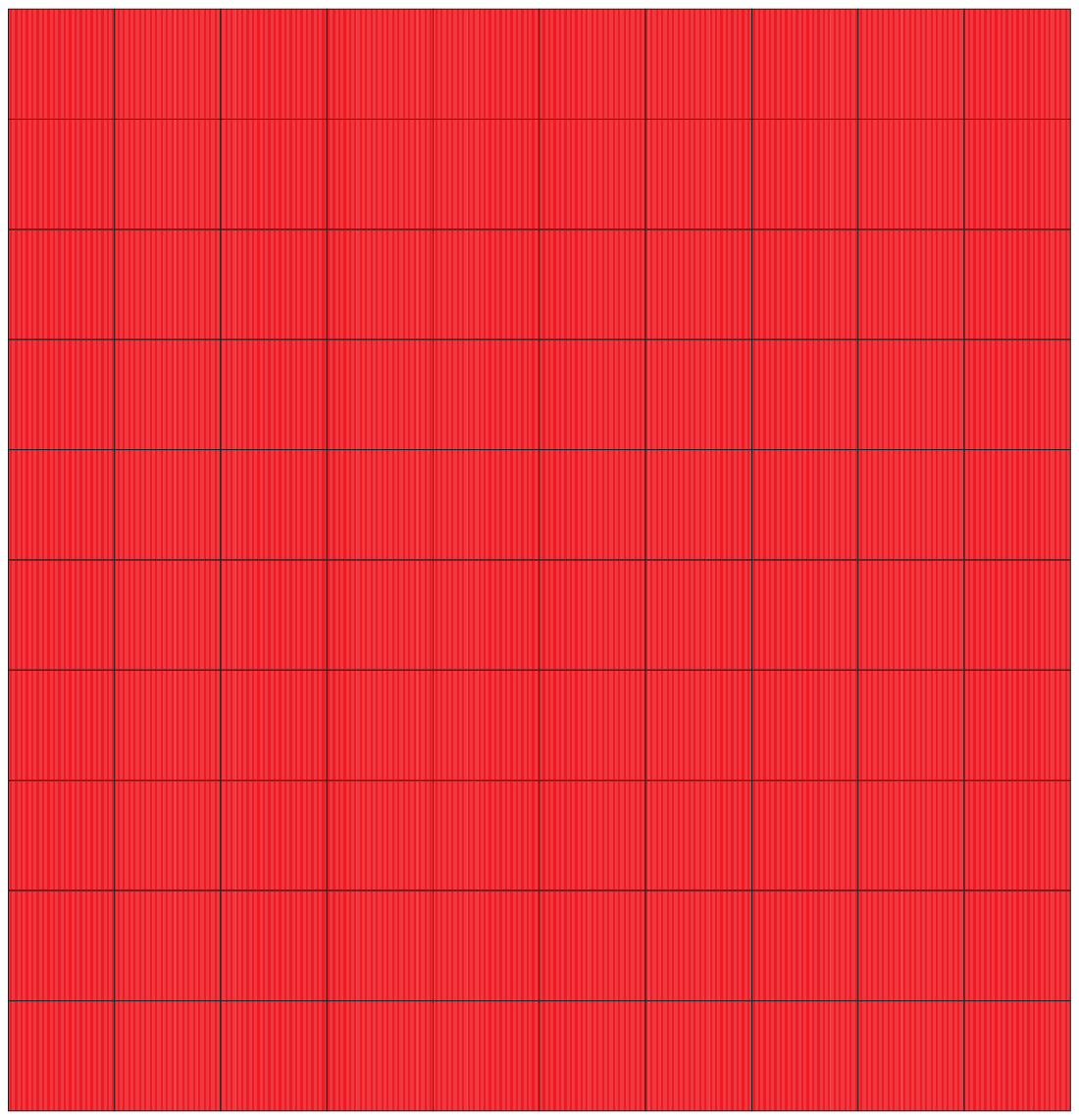}
  }
  \end{center}\vspace*{-2em}
\caption[]{(Color online) Time evolution of cooperator-invasion on a $10 \times 10$ square lattice with 8-player neighborhood. Initially, there are four cooperators (the four red squares in figure(a)) and ninety-six defectors (the ninety-six blank square in figure (a)). If $R < 3M/10$, then the cooperators take over the entire population in the stationary state ($t=5$, figure (d)).}
\label{fig:c8invade} \vspace*{-0.3cm}
\end{figure}

As observed in Fig.~\ref{fig:fcr}, the cooperation level decreases monotonically as $R$ increases until $R$ is greater than the threshold $R_t$ ($R_t \approx 40$).
To interpret this observation, here we study the relationship between the cooperation level of the system ($\rho _c$) and the parameter $R$ in an ideal model where the players on the square lattice can only select two pure strategies: $R$ or $M$.  First, we consider the case of the cooperator invasion. For simplicity, we assume that the system initially has four cooperators (players selecting strategy $M$) forming a square cluster, and all the other players are defectors (players selecting strategy $R$). Under such initial state, for the square lattice with 4-player neighborhood, we have the following results: (1) if $R < 2M/5$, the cooperators conquer the whole population, (2) if $R > 2M/5$, the cooperators are extinct, and (3) if $R = 2M/5$, cooperators and defectors are coexistent (the initial state is unchanged). Similarly, for the square lattice with 8-player neighborhood, we have the following results: (1) if $R < 3M/10$, the cooperators invade the whole population, (2) if $R > 3M/10$, the cooperators are extinct, and (3) if $R = 3M/10$, then cooperators and defectors are coexistent (the initial state is unchanged). Fig.~\ref{fig:c4invade} and Fig.~\ref{fig:c8invade} illustrate the time evolution of cooperator-invasion on a $10 \times 10$ square lattice with 4-player neighborhood and 8-player neighborhood respectively. As desired, if the conditions of the cooperator-invasion are satisfied, the cooperators take over the whole population in the stationary state as illustrated in Fig.~\ref{fig:c4invade} and Fig.~\ref{fig:c8invade}.

\begin{figure}[t]
\begin{center}
 \subfigure[t=1]{
     \includegraphics[width=0.2\columnwidth]{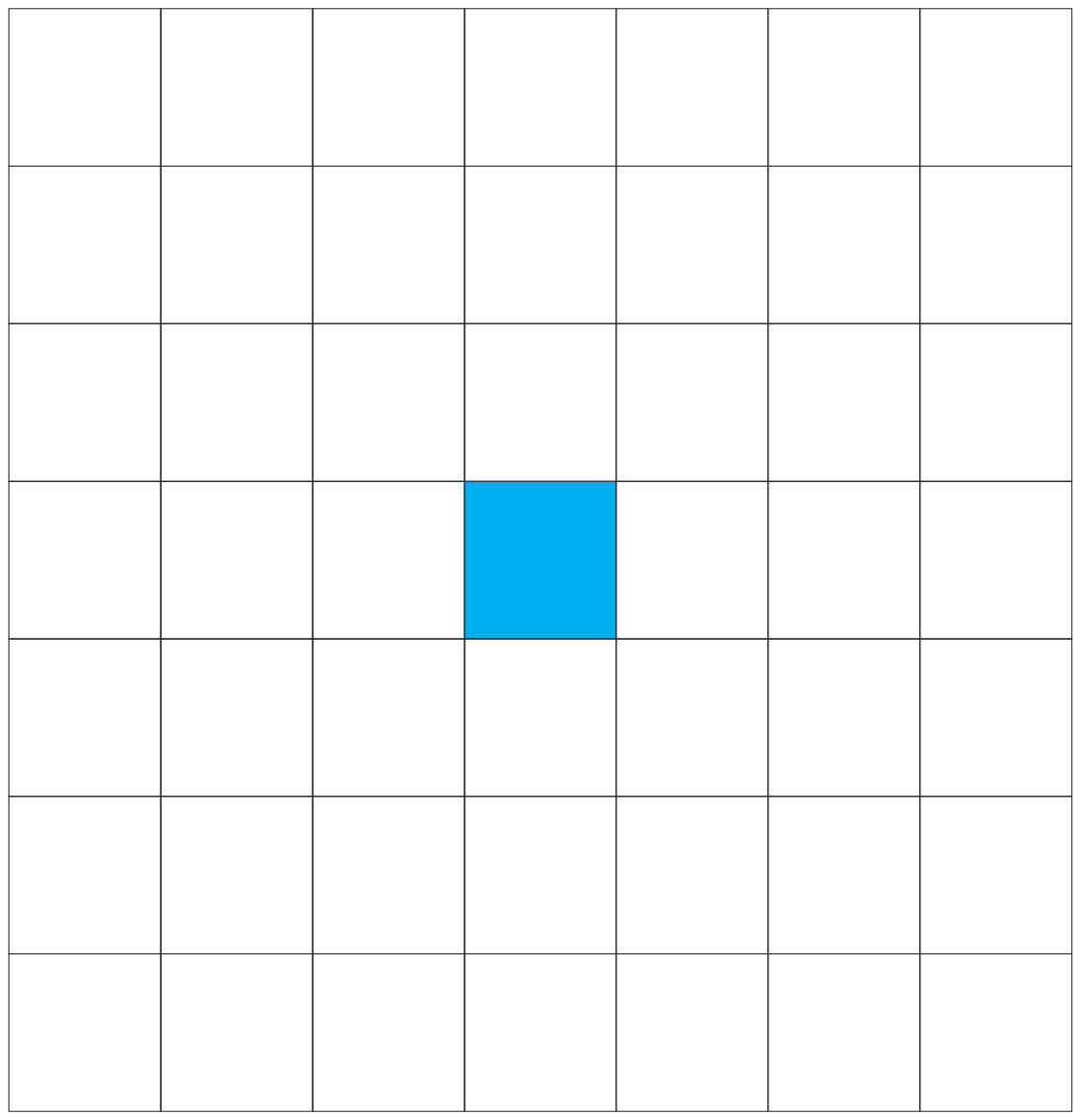}
 }
 \subfigure[t=3]{
     \includegraphics[width=0.2\columnwidth]{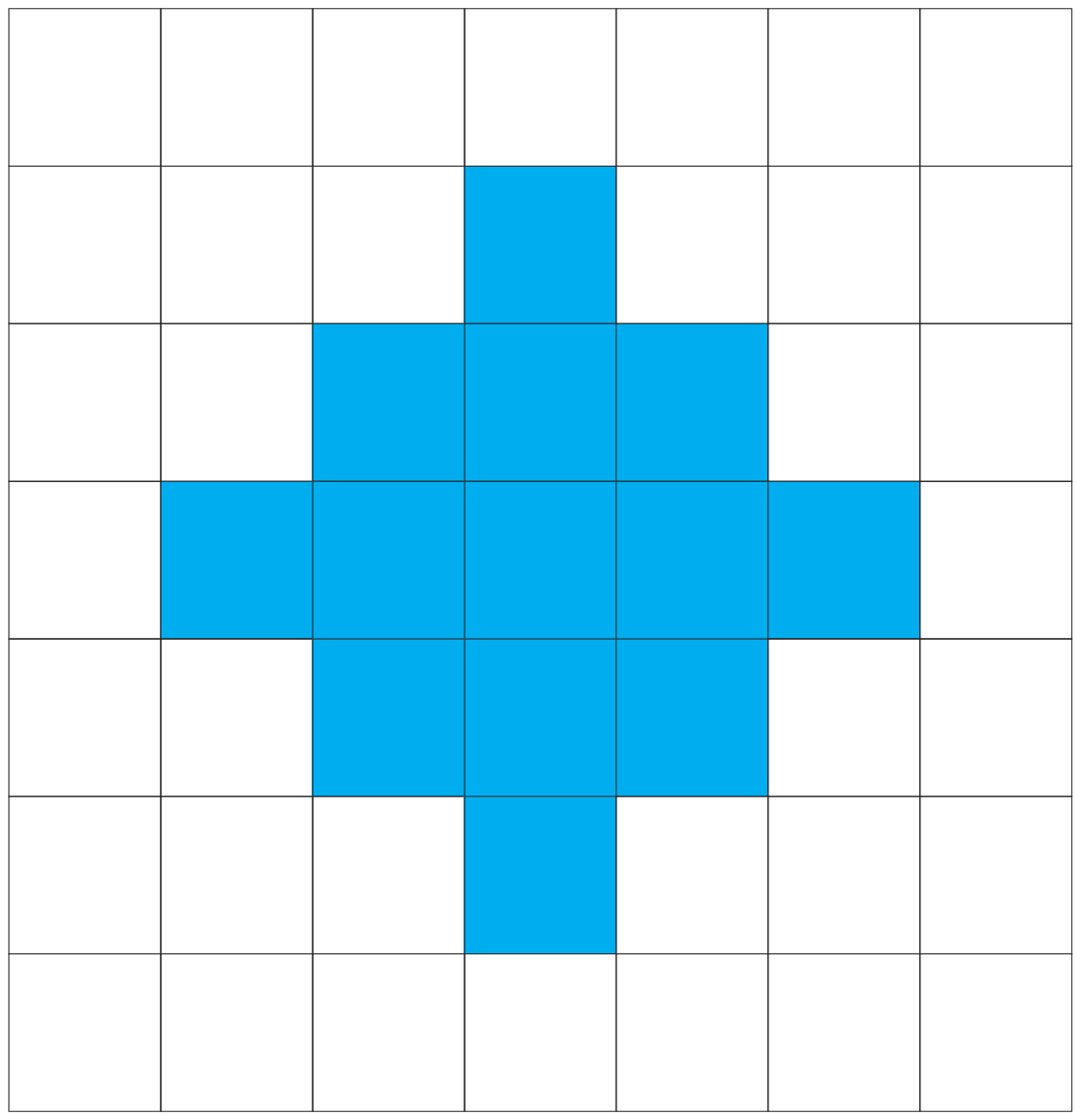}
  }
 \subfigure[t=5]{
     \includegraphics[width=0.2\columnwidth]{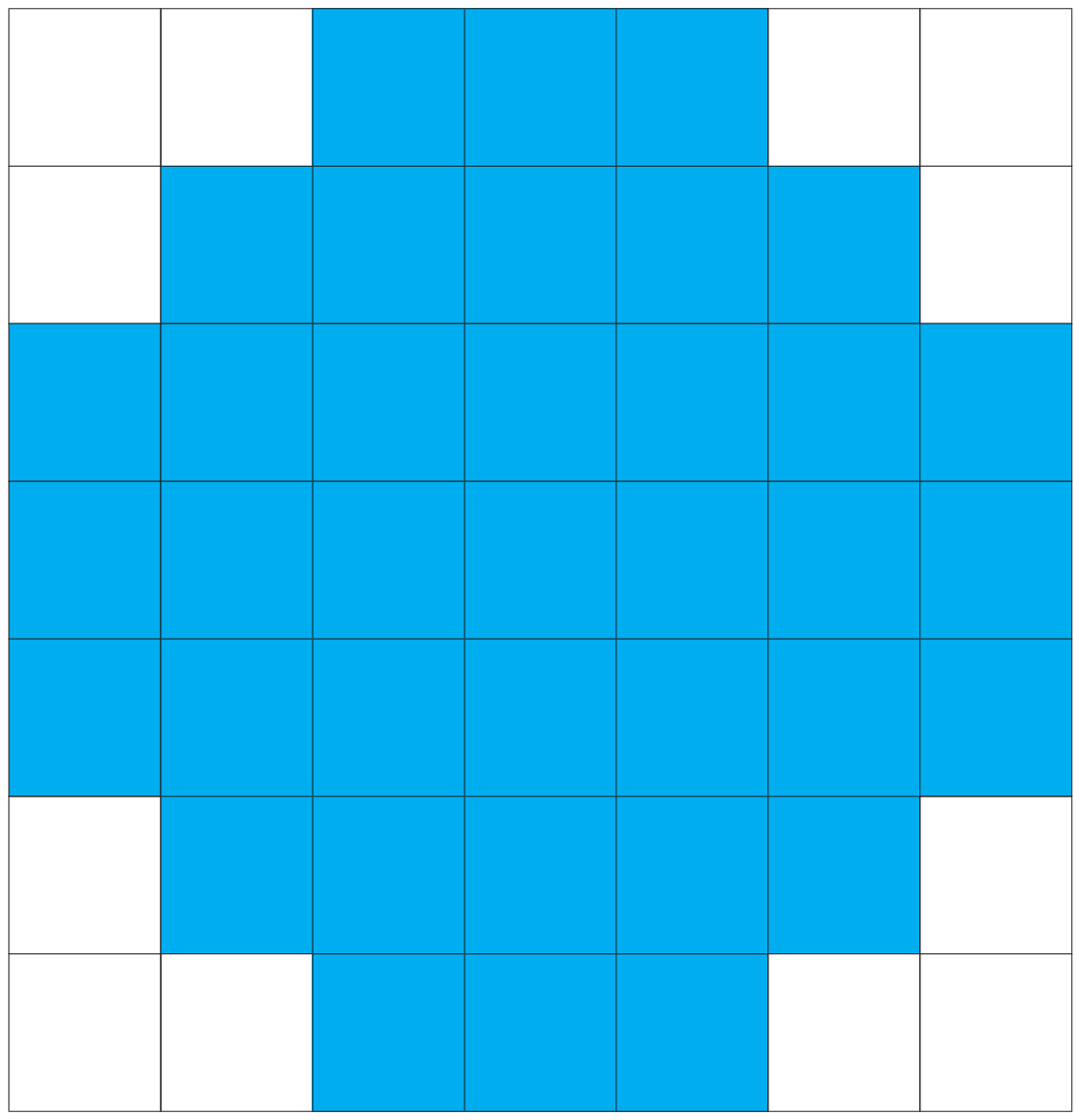}
  }
  \subfigure[t=7]{
     \includegraphics[width=0.2\columnwidth]{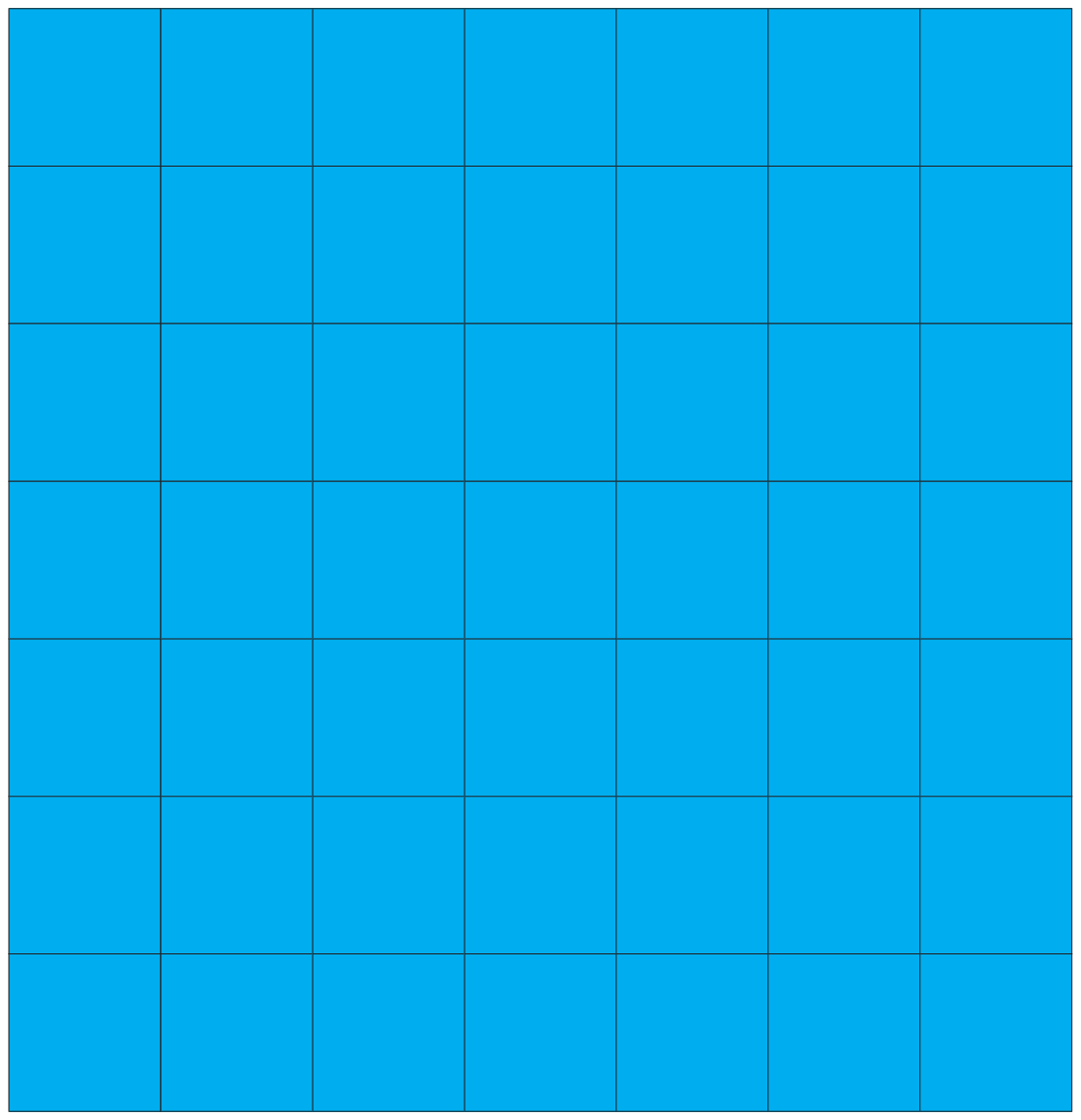}
  }
  \end{center}\vspace*{-2em}
\caption[]{(Color online) Time evolution of defector-invasion on a $7 \times 7$ square lattice with 4-player neighborhood. Initially, there is only one defector (the green square in figure (a)) and forty-six cooperators (the forty-six blank squares). If $R > 2M/3$, the defectors invade all the squares in the stationary state ($t=7$, figure(d)). }
\label{fig:d4invade} \vspace*{-0.3cm}
\end{figure}

\begin{figure}[t]
\begin{center}
 \subfigure[t=1]{
     \includegraphics[width=0.2\columnwidth]{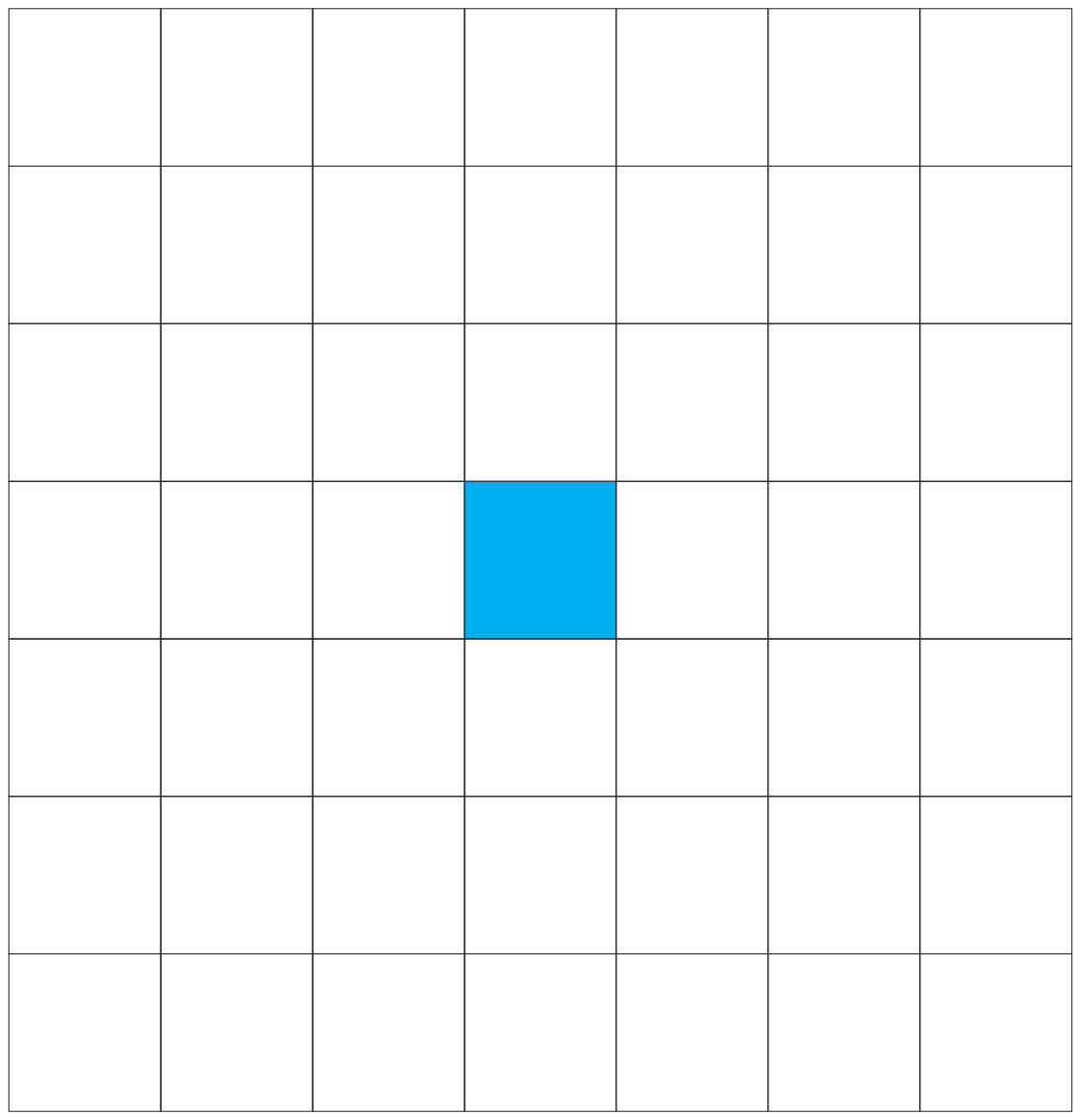}
 }
 \subfigure[t=2]{
     \includegraphics[width=0.2\columnwidth]{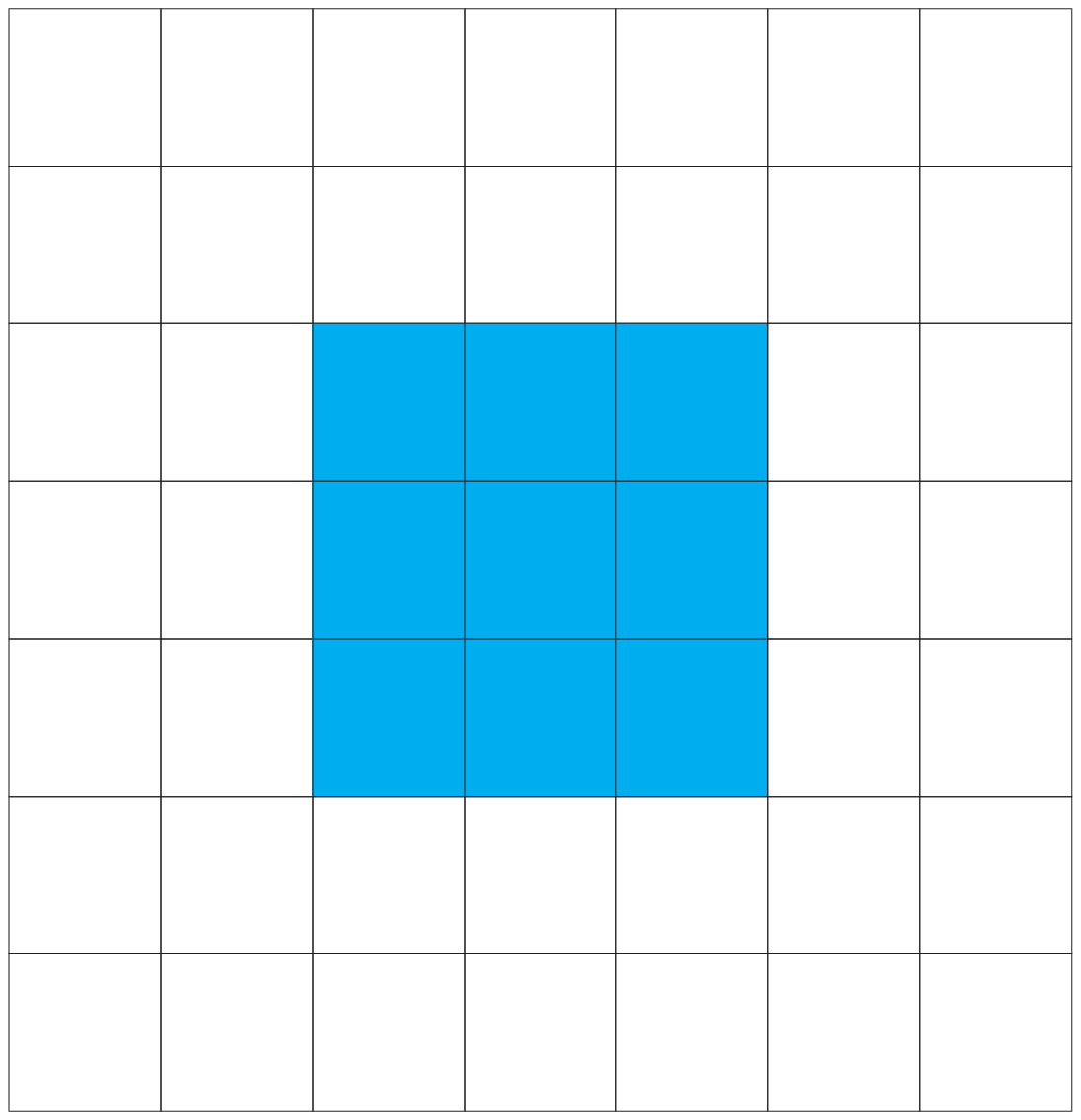}
  }
 \subfigure[t=3]{
     \includegraphics[width=0.2\columnwidth]{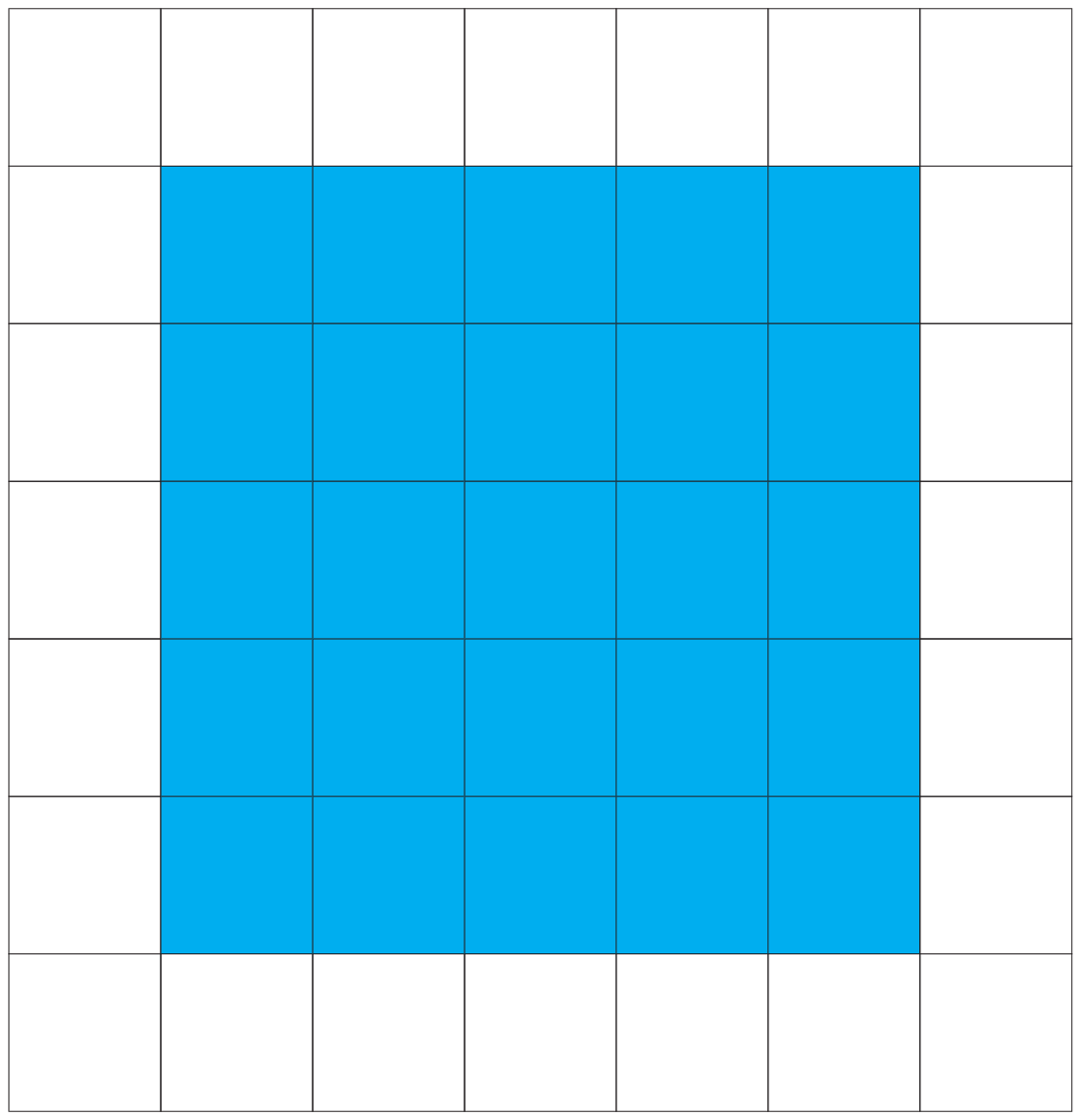}
  }
  \subfigure[t=4]{
     \includegraphics[width=0.2\columnwidth]{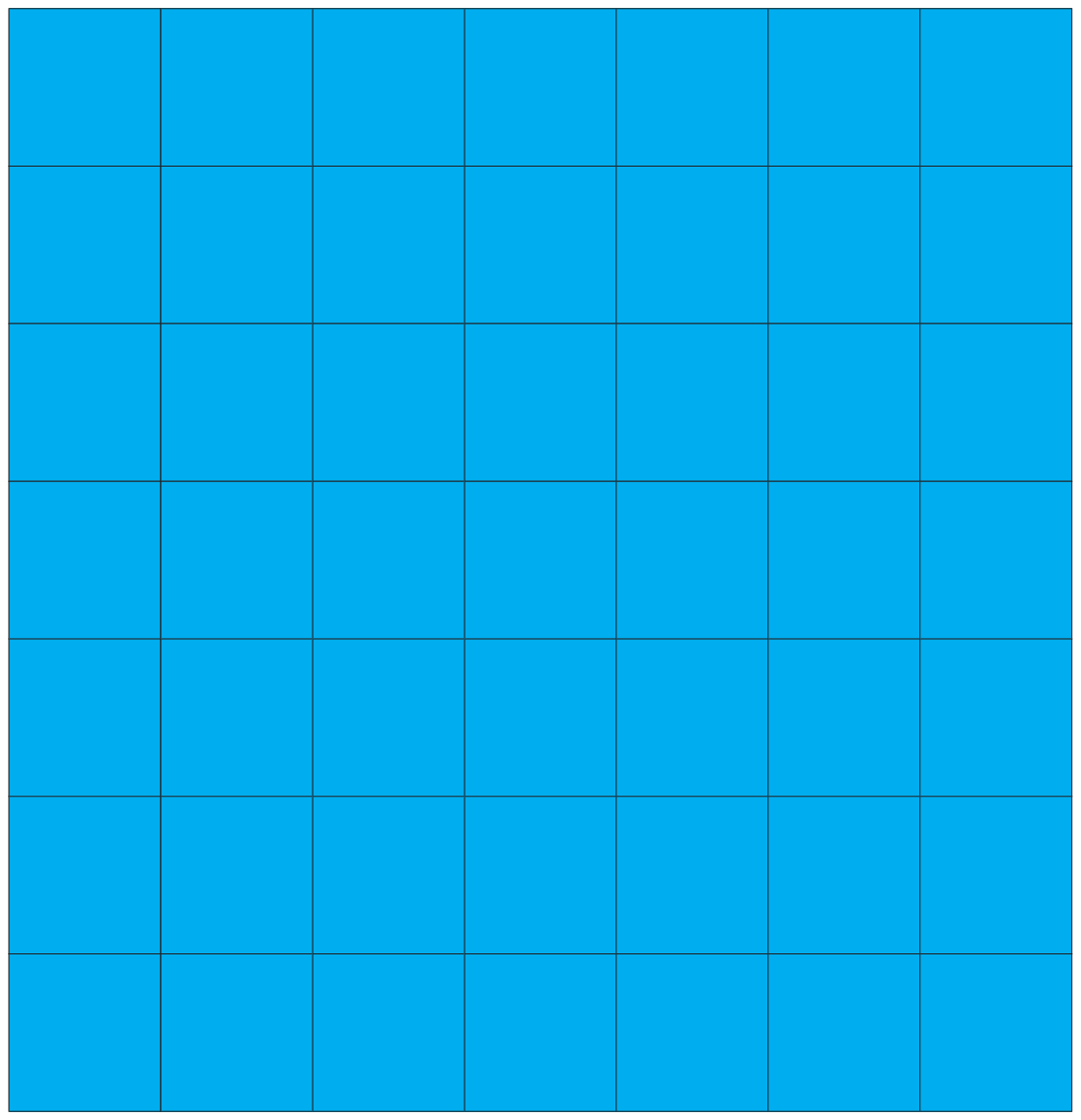}
  }
  \end{center}\vspace*{-2em}
\caption[]{(Color online) Time evolution of defector-invasion on a $7 \times 7$ square lattice with 8-player neighborhood. Initially, there is one defector (the green square in figure(a)) and forty-six cooperators (the forty-six blank squares). If $R > 8M/11$, the defectors occupy all the squares in the stationary state ($t=4$, figure (d)).}
\label{fig:d8invade} \vspace*{-0.3cm}
\end{figure}

Second, we consider the case of defector invasion. Suppose that the system initially has only one defector who selects strategy $R$, and the rest of the players are cooperators who select strategy $M$. Under such initial configuration, for the square lattice with 4-player neighborhood, we have the following results: (1) if $R < 3M/8$, the defectors vanish in the stationary state, (2) if $3M/8 \le R \le 2M/3$, the defectors and cooperators will be coexistent in the stationary state, and (3) if $R > 2M/3$, the defectors conquer the whole population in the stationary state. Likewise, for the square lattice with 8-player neighborhood, we can derive that (1) if $R < 7M/16$, defectors will disappear in the stationary state, (2) if $7M/16 \le R \le 8M/11$, the defectors and cooperators will coexist, and (3) if $R > 8M/11$, the defectors will take over the entire population. Fig.~\ref{fig:d4invade} and Fig.~\ref{fig:d8invade} depict the time evolution of defector-invasion on a $7 \times 7$ square lattice with 4-player neighborhood and 8-player neighborhood respectively. From Fig.~\ref{fig:d4invade} and Fig.~\ref{fig:d8invade}, we can clearly see that if the conditions of defector-invasion are met, then the defectors will occupy the whole lattice.

Based on our analysis in the ideal models, for the square lattice with 4-player neighborhood, we can conclude that, if $R < 2M/5$, then cooperator invasion will emerge and if $R > 2M/3$, then there is no cooperator in the system. For a large square lattice system, we can approximately analyze a small sub-lattice (eg. $10 \times 10$) by applying our results in the ideal model. In such a small sub-lattice, assume that four players who form a $2 \times 2$ cluster as illustrated in Fig.~\ref{fig:c4invade}(a) choose the strategy $M^\prime$ ($R < M^\prime  \le M$) and all the other players in the sub-lattice are defectors who choose the strategy $R$. If $R$ is large, then the cooperator-invasion condition (i.e., $R < 2M^\prime/5$) could not be met. As a result, all the cooperators would vanish, and thereby the defectors will occupy the small sub-lattice. Further, the sub-lattice occupied by the defectors would spread out over the whole lattice given a large $R$, thus resulting in a low cooperation level. In contrast, if $R$ is small, then the cooperator-invasion condition (i.e., $R < 2M^\prime/5$) could be satisfied, and thereby the cooperators invade the small sub-lattice, and then form a cooperator-cluster which can defend the invasion of defectors. If $R$ is small enough, the cooperator-cluster could spread out over the whole system, leading to a high cooperation level. On the other hand, suppose only one player selects the smallest strategy $R$ and all the other players in the sub-lattice select strategy $M ^\prime $ (i.e., $R < M^\prime  \le M$). If $R$ is large, the defector-invasion condition $R>2  M^\prime/3$ could be easily satisfied, thereby the sub-lattice could be occupied by the defectors. Then, the defectors form a cluster which could spread out over the whole lattice, thus resulting in a low $\rho_c$. On the contrary, if $R$ is small, then the defector-invasion condition (i.e., $R>2  M^\prime/3$) could not be satisfied. Moreover, if the condition $R < 3M^\prime/8$ is met, then the cooperators will occupy the small sub-lattice. Consequently, the cooperators will form a cooperator-cluster, and then they could spread out over the whole lattice, which leads to a high $\rho_c$. Put it all together, we conclude that large $R$ promotes defector invasion, while small $R$ facilitates cooperator invasion. Therefore, the cooperation level of the system ($\rho_c$) exhibits an inverse relationship with the parameter $R$. In addition, it is worth noting that if $R > 2M/3$ (implying $R>2  M^\prime/3$), then the system will be dominated by the defectors. Hence, the threshold of the system must be smaller than $2M/3$. Our result in Fig.~\ref{fig:fcr} (left panel) shows that the threshold is around 40, which is clearly smaller than $2M/3 \approx 67$. Similar analysis can be done in the square lattice with 8-player neighborhood.

\begin{figure}
\begin{center}
  \includegraphics[width=\columnwidth]{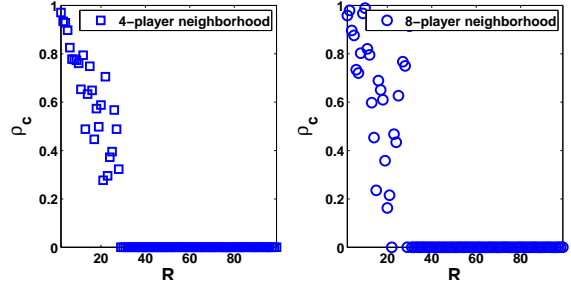}
\end{center}\vspace*{-2em} \caption[]{(Color online) Cooperation level $\rho_c$ as a function of the parameter $R$  on square lattices with 4-player neighborhood (left panel) and 8-player neighborhood (right panel) under the random strategy-adoption rule. All of the results are obtained at $\tau = 0.1$.} \label{fig:fcrrnd} \vspace*{-0.3cm}
\end{figure}

Now we turn to report the result of the spatial TD game with a random strategy-adoption rule. Fig.~\ref{fig:fcrrnd} depicts our results for $\rho_c$ as a function of $R$ on two lattice models at $\tau = 0.1$. Similar to the deterministic strategy-adoption case, we can observe that the cooperation emerges given $R$ is smaller than the threshold $R_t$, and $\rho_c$ decreases monotonically with increasing $R$ until the threshold $R_t$, where $\rho_c = 0$. Moreover, we find that the results for the model with the random strategy-adoption rule are robust to the noise parameter $\tau$ (not shown). There is a minor difference from the deterministic strategy-adoption case. The threshold of the model with random strategy-adoption is smaller than those of the model with deterministic strategy-adoption. These results could indicate that the deterministic strategy-adoption rule could be better than the random strategy-adoption rule to promote the emergence of cooperation in spatial TD game.

\section{Conclusions}\label{sec:concld}
To summarize, we have investigated the evolution of cooperation in spatial TD game, where the players are placed on a square lattice. An individual gains payoff by playing TD game with his immediate neighbors. Two evolutionary rules, namely deterministic strategy-adoption rule and random strategy-adoption rule are studied in the spatial TD game model. More specifically, for the deterministic strategy-adoption rule, each player revises the strategy based on his payoff and his neighbors' payoffs. For the random strategy-adoption rule, a randomly-selected player adopts one of his neighbors' strategies with a probability depending on the difference of their payoff. We apply Monte-Carlo simulation to our models, and the results show that the cooperation level of the spatial TD game has an inverse relationship with the parameter $R$. In particular, the cooperation level decreases monotonically with increasing $R$ until $R$ reaches the threshold $R_t$, where the cooperation level vanishes. By visualizing the spatial patterns of our models, we find that the cooperators who select the same large strategy will form clusters in the stationary state, and such clusters can resist the invasion of defectors. To further explain our findings, we analyze the conditions of both cooperator-invasion and defector-invasion in an ideal model, where the players are given two pure strategies to select: $R$ or $M$. Our analysis implies that the large $R$ hampers cooperator invasion and facilitates defector invasion, while the small $R$ promotes cooperator invasion and impedes defector invasion. As a result, the cooperation level of the system exhibits an inverse relationship with the parameter $R$.

Our findings suggest that the spatial reciprocity can promote the evolution of cooperation in TD game. Furthermore, these findings indicate that the spatial TD game model can be used to interpret the anomalous behavior in TD game that is observed in many previous behavioral studies \cite{99aertd, 99pnastd}. We hope that this work will inspire future studies on investigating the evolution of cooperation in spatial TD game, which has attracted little attention in physics community. For example, one promising direction is to study the impact of network structure on the evolution of cooperation in spatial TD game.

\begin{acknowledgements}
  The work was supported by a grant from the Research Grants Council of
  the Hong Kong SAR, China No. CUHK/419109.
\end{acknowledgements}

\bibliographystyle{apsrev4-1}
\bibliography{tdgame}
\end{document}